\documentclass[aps,prl,twocolumn,footinbib,bibnotes,preprintnumbers]{revtex4-1} 

\usepackage[utf8]{inputenc}
\usepackage{amsmath,amsfonts,epsfig,color,latexsym}
\usepackage{amssymb}
\usepackage{epsfig,psfrag}
\usepackage{slashed}
\usepackage[normalem]{ulem}

\usepackage{soul,xcolor}
\usepackage{hyperref}
\setstcolor{red}
\usepackage{comment}

\newcommand{\la}{\langle}
\newcommand{\ra}{\rangle}

\usepackage{tikz}

\def\cC{{\cal C}}
\def\ie{\begin{equation}\begin{aligned}}
\def\fe{\end{aligned}\end{equation}}

\renewcommand{\a}{\alpha}
\newcommand{\cO}{{\cal O}}

\newcommand{\cT}{{\cal T}}

\newcommand{\tr}{\mbox{tr}}

\begin{document}
\newpage
\pagenumbering{arabic}

\title{Non-planar integrated correlator in $\mathcal{N}=4$ SYM}

\author{Shun-Qing Zhang}
\email{sqzhang@mpp.mpg.de}
\affiliation{Max-Planck-Institut f\"{u}r Physik, Werner-Heisenberg-Institut, D-80805 M\"{u}nchen, Germany}

\preprint{MPP-2024-90}

\begin{abstract}
Integrated correlator of four superconformal stress-tensor primaries in $SU(N)$ $\mathcal{N}=4$ super Yang-Mills (SYM) theory in the perturbative limit takes a remarkably simple form, where the $L$-loop coefficient is given by a rational multiple of $\zeta(2L+1)$.
In this letter, we extend the previous analysis of expressing the perturbative integrated correlator as a linear combination of periods of $f$-graphs, graphical representations for loop integrands, to the non-planar sector at four loops. At this loop order, multiple zeta values make their first appearance when evaluating periods of non-planar $f$-graphs, but cancel non-trivially in the weighted sum. The remaining single zeta value, along with the rational number prefactor, makes a perfect agreement with the prediction from supersymmmetric localisation. 
\end{abstract}

\setcounter{tocdepth}{2}
\maketitle
\setcounter{page}{1}
\section{Introduction}
The exact results of integrated correlators for four stress-tensor operators in $SU(N)$ $\mathcal{N}=4$ SYM have been recently proposed in \cite{Dorigoni:2021bvj, Dorigoni:2021guq} for finite $g_{\rm YM}$ coupling and finite $N$ (see also the review \cite{Dorigoni:2022iem} and earlier works for large-$N$ \cite{Chester:2019jas, Chester:2020vyz}), based on the techniques from supersymmetric localisation \cite{Binder:2019jwn,Chester:2019pvm, Chester:2020dja}.
Having these exact results provides great insights into perturbative and non-perturbative physics, e.g. as shown in  \cite{Dorigoni:2021bvj, Dorigoni:2021guq,Dorigoni:2022zcr}, the weak coupling expansion of 
the integrated correlator (see \eqref{eq:d^2_m_pert}) exhibits an extremely simple pattern, where 
only \textit{single} zeta values show up at each loop order. This claim has been explicitly verified up to four loops in the planar limit in \cite{Wen:2022oky}, by making contact with periods of Feynman integrals whose integrands were constructed in \cite{Eden:2011we,Eden:2012tu} by graphical methods, so called $f$-graphs (see also \cite{Bourjaily:2015bpz,Bourjaily:2016evz} for integrands up to ten loops in the planar limit).
Unlike the planar sector, the non-planar part of physics is less explored. In view of that, we extend the construction in \cite{Wen:2022oky} to the non-planar sector at four loops, confirming the prediction from localisation by a direct computation of Feynman periods. The four-loop non-planar integrand was given in \cite{Eden:2012tu} with the coefficients fixed in \cite{Fleury:2019ydf}
\footnote{The same integrand has also been applied to determine the full four-loop cusp anomalous dimension in SYM \cite{Henn:2019swt}.}.
Despite this integrand is explicitly given, it still remains challenging to evaluate those periods at higher loop orders
\footnote{Here we are interested in periods for four-loop non-planar $f$-graphs, which are certain five-loop two-point Feynman integrals in position space \cite{Georgoudis:2018olj}, while they are still not fully known in the non-planar sector.}
. To circumvent  this, we found a particularly good choices \eqref{eq:c_4loops} by utilising Gram determinant conditions; as a result, all the difficult integrals are eliminated, with leftover ones easily being evaluated by the {\tt Maple} program {\tt HyperLogProcedures} \cite{HyperlogProcedures}. 
\section{Integrated correlator in $\mathcal{N}=4$ SYM}
The observable of interest is the four-point correlation with all four operators in the stress-tensor multiplet
\begin{align}
&\langle \cO_2(x_1, Y_1)\dots \cO_2(x_4, Y_4)  \rangle \nonumber\\
&\hspace{0.5cm}= {\rm \textit{free part.}} + \frac{1}{x_{12}^4 x_{34}^4}\mathcal{I}_4(U,V; Y_i) \cT(U,V) \,.
\end{align}
The weight-two half-BPS operator is defined as (see also the review \cite{Heslop:2022xgp})
\begin{equation}
    \cO_2(x, Y) :=  \tr(\Phi^{I}(x) \Phi^J(x)) Y_I Y_J \,,
\end{equation}
where $\Phi^{I}(x)$ are the six fundamental scalars in the $\mathcal{N}=4$ SYM theory contracting with
the null vectors $Y_I$.
The four-point function has been separated into the free and dynamic part, the latter takes a factorized form with all $Y_i$ dependence packaged in a well-known prefactor 
$\mathcal{I}_4(U,V; Y_i)$ that is fixed by the superconformal symmetry \cite{Eden:2000bk,Nirschl:2004pa}, and the four-point cross ratios are 
\begin{align} \label{eq:UV}
   &U=\frac{x_{12}^2x_{34}^2}{x_{13}^2x_{24}^2}\,, \; \qquad \quad V=\frac{x_{14}^2x_{23}^2}{x_{13}^2x_{24}^2} \, .
\end{align}

The integaretd correlator is defined as integrating $\cT(U,V)$ over spacetime coordinates $U$ and $V$, along with a specific measure \cite{Binder:2019jwn, Chester:2020dja} \footnote{In the literature, there exist two kinds of integrated correlators, while we focus on the first integrated correlators in this letter.}, which results in a function of  ’t Hooft couplings $\lambda=g^2_{\rm YM} N$ as the following
\begin{align} \label{eq:d^2_m}
\cC(\lambda) 
&:=I_2\left[\mathcal{T}(U,V)\right]\nonumber\\
&=-{8\over \pi} \int_0^{\infty} dr \int_0^{\pi} d\theta {r^3 \sin^2\theta \over U^2} \cT(U,V) \,,
\end{align}
where  and $r,\,\theta$ are linked to cross ratios as $U = 1+r^2 -2r \cos\theta$ and $V=r^2$.
As shown in \cite{Dorigoni:2021bvj,Dorigoni:2021guq}, the perturbative expansion of the integrated correlator 
in \eqref{eq:d^2_m}, i.e. small $g^2_{\rm YM}$ and finite $N$,
has the following form 
\begin{align}
\label{eq:d^2_m_pert}
&\cC^{pert}(\lambda)
\nonumber\\
&=4\,c\, \Big[ \frac{3   \, \zeta (3) a   }{2} -\frac{75 \, \zeta (5)a^2}{8} 
+\frac{735 \,\zeta (7) a^3}{16}  \nonumber\\
&\qquad\quad -\left(\frac{6615  \,\zeta (9)   } {32} + \mathbb{P}_{1}\right)a^4 + \mathcal{O}(a^{5}) \Big]\,,
\end{align}
where $c=\frac{N^2-1}{4}$ and $a={\lambda}/({4 \pi^2})$. The non-planar terms start to contribute at four loops
\begin{align} \label{eq:PGN}
\mathbb{P}_{1} = \frac{2}{7N^2} \times \frac{6615 \,\zeta (9) } {32}\,,
\end{align}
which we will later show is indeed the correct numerical factor to be consistent with the non-planar data in \cite{Fleury:2019ydf}.
\section{Perturbative integrated correlator as Feynman Periods}
To compute the integrated correlator in the weak coupling limit, we start with the dynamic part in the integrand \eqref{eq:d^2_m}, i.e.
the un-integrated correlator $\cT(U, V)$, 
which in the perturbative expansion is related to a familiar expression, $F^{(L)}(x_i)=F^{(L)}(x_1,x_2,x_3,x_4)$ in \cite{Eden:2011we,Eden:2012tu}, through the following
\begin{align}  \label{eq:TGN}
\cT(U, V) = 2\, c \,{U \over V} \sum_{L=1}^{\infty} a^L \, x_{13}^2 x_{24}^2\, F^{(L)}(x_i) \, .
\end{align}
In principle, one could plug the un-integrated correlator \eqref{eq:TGN} into \eqref{eq:d^2_m} to compute the integrated one; however, this will
involve complicated expressions of polylogarithms, and the analytical results are best known up to three loops \cite{Drummond:2013nda}.
To go beyond the three-loop order in \eqref{eq:d^2_m_pert}, 
an observation was made in \cite{Wen:2022oky} that the $\cC^{pert}(\lambda)$ 
are simply given by a linear combination of the periods of $f$-graphs, where $f$-graphs are provided up to ten loops in the planar limit \cite{ Bourjaily:2015bpz, Bourjaily:2016evz}. More importantly, the non-planar $f$-graphs at four loops are given in \cite{Eden:2012tu}, where the coefficients were later fixed by \cite{Fleury:2019ydf}.

In \cite{Wen:2022oky}, the perturbative integrated correlator are shown to be the following
\begin{align} \label{eq: I2_expansion}
&I_2\left[\mathcal{T}(U,V)\right] :=4\,c \sum_{L \geq 1} a^L I_2^{\prime} \left[ F^{(L)}(x_i)  \right]
\nonumber\\
&=-4\, c \sum_{L \geq 1} { a^L\over L! (-4)^L} \sum_{\alpha=1}^{n_L} c_{\alpha}^{(L)}\, \mathcal{P}_{f^{(L)}_{\alpha}}
\,,
\end{align} 
where the first equality is simply plugging \eqref{eq:TGN} into \eqref{eq:d^2_m_pert}, and the second equality makes use of the relation between $F^{(L)}(x_i)$ and $f^{(L)}(x_i)=f^{(L)}(x_1, x_2, \ldots, x_{4+L})$ as the following
\begin{equation} \label{eq:Ftof_function}
F^{(L)}(x_i)=\frac{\prod_{1\le i< j \le 4} x_{ij}^2}{L!(-4\pi^2)^L} \int d^4x_5 \cdots d^4x_{4+L} \,f^{(L)}(x_i)\,.
\end{equation}
The function $f^{(L)}(x_i)$ is written as a linear combination of $f^{(L)}_{\alpha}(x_i)$ with the subscript $\a$ denoting different topologies, and the coefficients are fixed by certain physical requirements \cite{Eden:2012tu,Bourjaily:2015bpz,Bourjaily:2016evz}, 
\begin{align}  \label{eq:f_sum}
f^{(L)}(x_i) = \sum_{\alpha=1}^{n_{L}} c_{\alpha}^{(L)} f^{(L)}_{\alpha}(x_1, x_2, \ldots, x_{4+L}) \,.
\end{align} 
Each function $f^{(L)}_\a(x_i)$, being totally symmetric under exchange of any pair of coordinates $x_i$ and $x_j$ due to hidden symmetry, is defined as \cite{Eden:2011we,Eden:2012tu}
\begin{equation}
f^{(L)}_{\alpha}(x_1, x_2, \ldots, x_{4+L}) = {P^{(L)}_{\alpha}(x_1, x_2, \ldots, x_{4+L}) \over \prod_{1 \leq i < j \leq 4+L} x_{ij}^2 } \,,
\end{equation}
where $P^{(L)}_\a$ is a homogeneous polynomials in $x^2_{ij}$ of degree $(L-1)(L+4)/2$, and it can be graphically determined by the so-called $P$-graphs \cite{Eden:2012tu}.
The period of $f^{(L)}_\a$  (see \cite{Broadhurst:1995km, Schnetz:2008mp, Brown:2009ta, Schnetz:2013hqa,  Panzer:2016snt, Panzer:2014caa, Borinsky:2020rqs} for more discussions on Feynman periods) is defined as the following
\begin{align}
\mathcal{P}_{f^{(L)}_{\alpha}}:=
\frac{1}{(\pi^2)^{L+1}}\int \frac{d^4 x_1\cdots d^4x_{4+L}}{{\rm vol} [{\rm SO(2,4)}]} f^{(L)}_{\alpha}(x_1, x_2, \ldots, x_{4+L}) \,.
\end{align}

Now we review the results of computing first three orders in \eqref{eq:d^2_m_pert} by using standard field theory techniques \eqref{eq: I2_expansion} \cite{Wen:2022oky}. 
Note that the function $f^{(L)}_\a$ at the first three loop order has only one planar topology, therefore we omit the subscript $\a$ for $L \leq 3$,
\begin{align} \label{eq:f_first_three_loops}
f^{(1)}(x_i)&=\frac{1}{\prod_{1\le i< j \le 5} x_{ij}^2}\,,\nonumber\\
f^{(2)}(x_i)&=\frac{\frac{1}{48}x_{12}^2 x_{34}^2 x_{56}^2}{\prod_{1\le i< j \le 6} x_{ij}^2}+S_6\,,
\nonumber\\
f^{(3)}(x_i)&=\frac{\frac{1}{20} x_{12}^4 x_{34}^2 x_{45}^2 x_{56}^2 x_{67}^2 x_{37}^2}{\prod_{1\le i< j \le 7} x_{ij}^2}+{S}_7\,,
\end{align}
where the numeric factors in the numerators, i.e. the $P^{(L)}$ polynomials, are to mod out the trivial $S_{4+L}$ permutations.
The periods of the above three $f^{(L)}$ functions can be easily computed by the {\tt Maple} programs {\tt HyperLogProcedures} \cite{HyperlogProcedures} and {\tt HyperInt} \cite{Panzer:2014caa}, which we give in the appendix \eqref{eq:periods_3loops}.

The integarted correlator expanded at the first three loop orders can be obtained by using \eqref{eq: I2_expansion} and \eqref{eq:periods_3loops} as
\begin{align} \label{eq:I2_from_periods_three_loops}
&I^{\prime}_2\left[{F^{(1)}}(x_i)\right]=-\frac{1}{1!(-4)^1} \times \mathcal{P}_{f^{(1)}} =\frac{3\zeta(3)}{2}\,,\nonumber\\
&I^{\prime}_2\left[{F^{(2)}}(x_i)\right]=-\frac{1}{2!(-4)^2} \times \mathcal{P}_{f^{(2)}}={-}\frac{75\zeta(5)}{8} \
\,,\nonumber\\
&I^{\prime}_2\left[{F^{(3)}}(x_i)\right]=-\frac{1}{3!(-4)^3} \times \mathcal{P}_{f^{(3)}}=\frac{735\zeta(7)}{16} \,.
\end{align}
The above results up to three loops are in total agreement with supersymmetric localisation \eqref{eq:d^2_m_pert} as demonstrated in \cite{Wen:2022oky}.
We stress again the simplicity of using period to compute pertubative integrated correlator, e.g. the $L=3$ result in \eqref{eq:I2_from_periods_three_loops} is a single-line computation using $\mathcal{P}_{f^{(3)}}$ (with specialist packages for periods \cite{HyperlogProcedures,Panzer:2014caa}). In contrast, using the original prescription \eqref{eq:d^2_m} will inevitably involve complicated expression of $F^{(3)}$ \cite{Drummond:2013nda}, which makes the computation infeasible.
\section{Four-loop integrated correlator}
As pointed out in \cite{Eden:2012tu}, starting at four loops, the loop integrand and corresponding $f$-function split into planar and non-planar parts
\begin{equation}
    f^{(4)}(x_i)=f^{(4)}_{g=0} (x_i)+ \frac{1}{N^2}\,f^{(4)}_{g=1} (x_i)\,,
\end{equation}
where $f^{(4)}_{g=0}$ consists of planar $f$-graphs only, while
$f^{(4)}_{g=1}$ includes both planar and non-planar $f$-graphs, i.e. $f$-graphs with genus either $0$ or $1$. 
In the following subsections, we will discuss the
integrated correlator at $L=4$ in a separated form 
\begin{align}
    I^{\prime}_2\left[{F^{(4)}} (x_i)\right]= I^{\prime}_2\left[{F^{(4)}_{g=0}} (x_i)\right]+\frac{1}{N^2}\,I^{\prime}_2\left[{F^{(4)}_{g=1}} (x_i)\right]\,,
\end{align}
where the planar and  non-planar contribution are obtained the summing periods of $f^{(4)}_{g=0}$ and $f^{(4)}_{g=1}$, respectively.
\subsection{Planar sector: periods of $f^{(4)}_{g=0}$}
The planar four-loop correlator is expressed as
sum of three topologies \cite{Eden:2012tu}
\begin{align}
    f^{(4)}_{g=0} (x_i)&=\sum_{\alpha=1}^3 c_{0;\alpha}^{(4)} f^{(4)}_{\alpha}(x_1,\cdots,x_8)\nonumber\\&=\sum_{\alpha=1}^3\,c_{0;\alpha}^{(4)}\frac{P_{\alpha}^{(4)}(x_1, \dots, x_8)}{\prod_{1\le i< j \le 8} x_{ij}^2}\,,
\end{align}
where the list of 3 coefficients is
\begin{equation}
    c^{(4)}_{0;\a}=\{1,1,-1\}\,,
\end{equation}
and the numerators $P_{\alpha}^{(4)}(x_i)$ are given by
\begin{align}  \label{eq:P-4loop-planar}
P^{(4)}_1(x_i)&=\frac{1}{24} \,x_{12}^2 x_{13}^2 x_{16}^2 x_{23}^2 x_{25}^2 x_{34}^2 x_{45}^2 x_{46}^2 x_{56}^2 x_{78}^6+S_8\,, \nonumber\\
P^{(4)}_2(x_i)&=\,\frac{1}{8}\;\, x_{12}^2 x_{13}^2 x_{16}^2  x_{24}^2 x_{27}^2 x_{34}^2 x_{38}^2 x_{45}^2 x_{56}^4 x_{78}^4+S_8\,,\nonumber \\ 
P^{(4)}_3(x_i)&=\frac{1}{16}\, x_{12}^2 x_{15}^2 x_{18}^2 x_{23}^2 x_{26}^2 x_{34}^2
   x_{37}^2 x_{45}^2 x_{48}^2 x_{56}^2 x_{67}^2 x_{78}^2\nonumber\\
   &\hspace{0.5cm}+S_8\,.
 \end{align}
According to \eqref{eq: I2_expansion}, the integrated correlator at four loops (planar sector) is then given by
\begin{align} \label{eq:planar_4loops}
    I^{\prime}_2\left[{F^{(4)}_{g=0}} (x_i)\right]&=-\frac{1}{4!\,(-4)^4} \times \left( \mathcal{P}_{f^{(4)}_1} +\mathcal{P}_{f^{(4)}_2} -\mathcal{P}_{f^{(4)}_3}\right) \nonumber\\
    &=-\frac{6615\zeta(9)}{32} \, , 
\end{align}
where we have used the periods of $f^{(4)}_\alpha$ given as the following
\begin{align} \label{eq:periods_four_loops_P}
     &\mathcal{P}_{f^{(4)}_1}= 8!\times \frac{1}{24} \times 252\zeta(9) \,, \nonumber \\
    & \mathcal{P}_{f^{(4)}_2}= 8!\times \;\frac{1}{8} \,\times 252\zeta(9) \,, \nonumber \\
     &\mathcal{P}_{f^{(4)}_3}= 8!\times \frac{1}{16} \times 168\zeta(9)\,.
\end{align}
The result of planar part \eqref{eq:planar_4loops} agrees with supersymmetric localisation \eqref{eq:d^2_m_pert} as shown in \cite{Wen:2022oky}.
\vspace{0.25cm}
\subsection{Non-planar sector: periods of $f^{(4)}_{g=1}$}
\vspace{-0.5cm}
The non-planar part of four-loop correlator consists of 32 topologies, including the first 3 planar ones in \eqref{eq:P-4loop-planar}, which can be expressed as
\begin{align} \label{eq:f4_NP}
    f^{(4)}_{g=1} (x_i)&=\sum_{\alpha=1}^{32} c_{1;\alpha}^{(4)} \,f^{(4)}_{\alpha}(x_1,\cdots,x_8)\nonumber\\&=\sum_{\alpha=1}^{32}\,c_{1;\alpha}^{(4)}\frac{P_{\alpha}^{(4)}(x_1, \dots, x_8)}{\prod_{1\le i< j \le 8} x_{ij}^2}\,,
\end{align}
where the 32 polynomials, $P_{\alpha}^{(4)}$, are defined in (C.1) of \cite{Eden:2012tu}. 

As mentioned in the introduction, the original non-planar data provided in \cite{Fleury:2019ydf} involves integrals that are hard to evaluate; to resolve this, we have chosen an alternative set of coefficients that is \textit{equivalent} to the one in the reference (the validity will be justified shortly using Gram determinant conditions). We choose the set of coefficients to be the following
\begin{align} \label{eq:c_4loops}
    c_{1;\alpha}^{(4)}=2\times \{&12,10,-14,8,-4,6,0,-1,-4,0,4,-2,-1,0^5,\nonumber\\
    &4,-2,4,0,-2,0^2,-2,0^6\} \,,
\end{align}
where a shorthand notation is adopted to express a list of $k$ zeros as $0^k$.
The periods of $f^{(4)}_{\alpha}$ that contributes to \eqref{eq:f4_NP}, i.e. with non-zero $c_{1;\alpha}^{(4)}$, are given in \eqref{eq:periods_four_loops_P} and \eqref{eq:periods_four_loops_NP} in the appendix.

As mentioned earlier, our choice of coefficients $c_{1;\alpha}^{(4)}$ differs from the one given in \cite{Fleury:2019ydf}, the equation (3.2) therein (the JHEP version) is 
\begin{align}
    2\tilde{q}
    =2\times \{&6,6,-6,8,0,6,0,-1,-2,0^2,2,-1,0^4,\nonumber\\
    &2,2,-2,-4,0,-2,0^3,-48,-4,0,4,0^2\}\,.
\end{align}
The coefficients $c_{1;\alpha}^{(4)}$ and $\tilde{q}$ are related by adding Gram polynomials that vanish in strictly four dimensions, i.e.
\begin{equation}
    0=\sum_{\a=1}^{32} a_{k,\a}\, P_{\alpha}^{(4)}(x_1, \dots, x_8) \,, \quad {\rm for} \;k=1,2,3\,,
\end{equation}
where $a_{k,\a}$ are three sets of 32 coefficients given as 
\footnote{Note the typos in (5.21) of \cite{Eden:2012tu}, where the first two entries of all $a_k$ therein should have been swapped.}
\begin{widetext}
\begin{align}
    a_1&=\{6,16,-8,8,-10,24,0,-4,8,6,-2,4,-4,-6,3,-9,0,3,4,-5,-2,-18,-2,3,-3,1,0^6\} \,,\nonumber\\
    a_2&=\{-9,-18,12,-8,12,-24,0,4,-7,-6,0,-2,4,6,-3,9,0,-2,-5,5,-2,18,2,-3,3,0,{-}24,-2,0,2,0^2\} \,,\nonumber\\
    a_3&=\{-1,4,2,4,-2,12,0,-2,5,2,-4,4,-2,-2,1,-3,-1,2,1,-1,-6,-6,0,1,0^2,-36,0,1,0,-1,1\}\,.
\end{align}
\end{widetext}
One can easily check
\begin{align}
&\sum_{\alpha=1}^{32}\,c_{1;\alpha}^{(4)}P_{\alpha}^{(4)}(x_1, \dots, x_8) \nonumber\\
    =&\sum_{\alpha=1}^{32}\,\left(2\tilde{q}-4\, (a_{1,\a}+a_{2,\a}) \right)P_{\alpha}^{(4)}(x_1, \dots, x_8)\,.
\end{align}
With the good choice of $\,c_{1;\alpha}^{(4)}$ in \eqref{eq:c_4loops} (instead of $\tilde{q}$), all the periods of $f^{(4)}_\a$ in \eqref{eq:f4_NP} with non-zero coefficients
can be directly evaluated by {\tt HyperLogProcedures} \cite{HyperlogProcedures}, and the result of the list of periods are given in \eqref{eq:periods_four_loops_NP} in the appendix.  

Finally, using expressions \eqref{eq:f4_NP}, \eqref{eq:c_4loops} and values of periods \eqref{eq:periods_four_loops_P}, \eqref{eq:periods_four_loops_NP}, the integrated correlator at four loops for the non-planar sector is given as
\begin{align} \label{eq:4loops_NP}
    I^{\prime}_2\left[{F^{(4)}_{g=1}} (x_i)\right]&=-\frac{1}{4!\,(-4)^4} \times \frac{1}{N^2}\times  \sum_{\a=1}^{32}\,c_{1;\alpha}^{(4)}\,\mathcal{P}_{f^{(4)}_\a}\nonumber\\
    &= -\frac{2}{7N^2}\times \frac{6615\zeta(9)}{32} \, , 
\end{align}
which, together with the planar part \eqref{eq:planar_4loops}, perfectly match the result from supersymmetric localisation \eqref{eq:d^2_m_pert}. In particular, periods for the four loop non-planar $f$-graphs $\mathcal{P}_{f^{(4)}_\a}$ contain different zeta values, including multi-zeta values such as $\zeta(5,3)$ in $\mathcal{P}_{f^{(4)}_4}$ and $\mathcal{P}_{f^{(4)}_{12}}$ in\eqref{eq:periods_four_loops_NP}
\begin{align*}
    &\mathcal{P}_{f^{(4)}_4}= 8! \times \frac{1}{16} \times\left( \frac{432}{5} {{\zeta(5,3)}}+252 \zeta(5) \zeta(3)-\frac{58 \pi ^8}{2625}\right) \,, \nonumber\\
    &\mathcal{P}_{f^{(4)}_{12}}=8! \times\,\frac{1}{4} \times\; \left( \frac{432}{5} {{\zeta(5,3)}}-36 \zeta(3)^2+360 \zeta(5) \zeta(3)\right.\\
    &\hspace{3cm}\left.+\frac{189 \zeta(7)}{2}-\frac{131 \zeta(9)}{2}-\frac{58 \pi ^8}{2625}
    \right) \,,
\end{align*}
while the $\zeta(5,3)$ parts above cancel out since $c_{1;4}=8,\,c_{1;12}=-2$, and the remaining products of zeta values will further cancel in the the linear combination in \eqref{eq:4loops_NP}.
\section{Summary and Outlook}
In this letter, we perform a first principle calculation of perturbative integrated correlators in terms of Feynman periods, with a focus on the non-planar sector at four loops \eqref{eq:4loops_NP}, which confirms the prediction from supersymmetic localisation \eqref{eq:d^2_m_pert}.
It is natural to consider second type of integrated correlator with a different measure \cite{Chester:2020dja,Chester:2020vyz,Alday:2023pet},
where some results (up to first three loops) have been investigated in \cite{Wen:2022oky}, which also displays a simple pattern of zeta value at each loop order.
Furthermore, it will be interesting to study integrated correlators involving more generic weights, such as
$\la 22pp\ra$ in \cite{Paul:2022piq,Brown:2023cpz,Brown:2023why}, 
and $\la p_1 p_2 p_3 p_4\ra$ in \cite{Brown:2023zbr} by utilising ten-dimensional (10D) conformal symmetry \cite{Caron-Huot:2021usw}, 
with possibilities to extend the non-planar limit. It will be fascinating to consider other types of integrated correlators, such as those involving a Wilson line \cite{Pufu:2023vwo,Billo:2023ncz}, or determinant operators \cite{Jiang:2023uut,Brown:2024tru}.
It is also worth mentioning the potential extension to $\mathcal{N}=2$ SYM, where the integrated correlators have been studied in \cite{Billo:2023kak,Pini:2024uia}.
Finally, it will be desirable to have an explanation of the simplicity of perturbative integrated correlator, i.e. only single zeta values allowed \eqref{eq:d^2_m_pert}, from a pure mathematical point of view; for example, the properties of periods and Galois coaction could play an important role \cite{brown2017feynman,Panzer:2016snt,Caron-Huot:2019bsq,Gurdogan:2020ppd}.
\section*{Acknowledgments}
We would like to thank Michael Borinsky, Song He, Johannes Henn, William Torres Bobadilla, and Chenyu Wang for many useful conversations.
We would especially thank Congkao Wen for insightful discussions and comments on the draft.
This research received funding from the European Research Council (ERC) under the European Union's Horizon 2020 research and innovation programme (grant agreement No 725110), {\it Novel structures in scattering amplitudes}.
\bibliographystyle{apsrev4-1} 
\bibliography{correlator.bib}

\onecolumngrid
\appendix
\section{Periods of $f$-graphs up to three loops}
Here we give the periods of $f^{(L)}_{\alpha}$ for $L\leq3$ computed by {\tt HyperLogProcedures} \cite{HyperlogProcedures} and {\tt HyperInt} \cite{Panzer:2014caa}, which can be applied to evaluate the perturbative integrated correlator at the first three orders \eqref{eq:I2_from_periods_three_loops}
\begin{align} \label{eq:periods_3loops}
     \mathcal{P}_{f^{(1)}}&=5!\times\frac{1}{120}\times\; 6 \zeta(3)\,, \nonumber\\
      \mathcal{P}_{f^{(2)}}&=6!\times\;\frac{1}{48}
      \,\times  20 \zeta(5)\,, \nonumber \\
       \mathcal{P}_{f^{(3)}}&=7!\times\,\frac{1}{20}\;\times 70 \zeta(7)\,,
\end{align}
where the $(4+L)!$ factor in each $\mathcal{P}_{f^{(L)}}$ is due to the $S_{4+L}$ permuatations in \eqref{eq:f_first_three_loops}, which gives  the same value of period for a given topology. 
\section{Periods of non-planar $f$-graphs at four loops} 
We present the periods of $f^{(4)}_{\alpha}$ that contributes to \eqref{eq:f4_NP}, see also the periods for planar $f^{(4)}_{\alpha}$ (where $\a=1,2,3$) presented in \eqref{eq:periods_four_loops_P}.
\begin{align} \label{eq:periods_four_loops_NP}
    \mathcal{P}_{f^{(4)}_4}&=8! \times \frac{1}{16} \times\left( \frac{432}{5} \zeta(5,3)+252 \zeta(5) \zeta(3)-\frac{58 \pi ^8}{2625}\right) \,, \nonumber\\
    \mathcal{P}_{f^{(4)}_5}&=8! \times \frac{1}{4}\times \left( 8 \zeta(3)^3+\frac{1063 \zeta(9)}{9}
    \right) \,, \nonumber\\
    \mathcal{P}_{f^{(4)}_6}&=8! \times \frac{1}{12}\times\left( 120 \zeta(5) \zeta(3)
    \right) \,, \nonumber\\
    \mathcal{P}_{f^{(4)}_8}&=8! \times \frac{1}{2}\times\left( 8 \zeta(3)^3+\frac{1567 \zeta(9)}{9}
    \right) \,, \nonumber\\
    \mathcal{P}_{f^{(4)}_9}&=8! \times \frac{1}{4}\times\left( 168 \zeta(9)
    \right) \,, \nonumber\\
    \mathcal{P}_{f^{(4)}_{11}}&=8! \times \frac{1}{4} \times \left( -36 \zeta(3)^2+108 \zeta(5) \zeta(3)+\frac{189 \zeta(7)}{2}
    \right) \,, \nonumber\\
    \mathcal{P}_{f^{(4)}_{12}}&=8! \times \frac{1}{4} \times \left( \frac{432}{5} \zeta(5,3)-36 \zeta(3)^2+360 \zeta(5) \zeta(3)+\frac{189 \zeta(7)}{2}-\frac{131 \zeta(9)}{2}-\frac{58 \pi ^8}{2625}
    \right) \,, \nonumber\\
    \mathcal{P}_{f^{(4)}_{13}}&=8! \times \frac{1}{2} \times \left( -24 \zeta(3)^3+120 \zeta(5) \zeta(3)+\frac{727 \zeta(9)}{6}
    \right) \,, \nonumber\\
    \mathcal{P}_{f^{(4)}_{19}}&=8! \times \frac{1}{4} \times \left(16 \zeta(3)^3+72 \zeta(3)^2+24 \zeta(5) \zeta(3)-189 \zeta(7)+\frac{2126 \zeta(9)}{9} \right) \,, \nonumber\\
    \mathcal{P}_{f^{(4)}_{20}}&=8! \times \frac{1}{4} \times \left( -16 \zeta(3)^3+72 \zeta(3)^2+144 \zeta(5) \zeta(3)-189 \zeta(7)+\frac{1906 \zeta(9)}{9} \right) \,, \nonumber\\
    \mathcal{P}_{f^{(4)}_{21}}&=8! \times \frac{1}{8} \times \left(120 \zeta(5) \zeta(3)\right) \,, \nonumber\\
    \mathcal{P}_{f^{(4)}_{23}}&=8! \times \frac{1}{8} \times \left(48 \zeta(3)^3-72 \zeta(3)^2+216 \zeta(5) \zeta(3)+189 \zeta(7)-\frac{388 \zeta(9)}{3}\right) \,, \nonumber\\
    \mathcal{P}_{f^{(4)}_{26}}&=8! \times \frac{1}{16} \times \left( 96 \zeta(3)^3+288 \zeta(3)^2+96 \zeta(5) \zeta(3)-756 \zeta(7)+\frac{1228 \zeta(9)}{3} \right) \,.
\end{align}
The multiple zeta value is defined by
\begin{equation}
    \zeta(n_d,\cdots,n_1)=\sum_{k_d>\cdots>k_1\geq 1} \frac{1}{k_d^{n_d}\,\cdots\,k_1^{n_1}}\,  ,\quad\quad n_d\geq 2\,. 
\end{equation}

\end{document}